\documentclass{icrc2009}

\usepackage{graphicx}   % for including figures
\usepackage[caption=false]{caption}    % for captions
\usepackage[font=footnotesize]{subfig} % subfig.sty for a double column floating figure using two subfigures
\usepackage{fixltx2e}
\usepackage{url}

\newcommand{\shorttitle}[1]%
{\markboth{Proceedings of the 31\MakeLowercase{$^{st}$} ICRC, {\L}\'{o}d\'{z} 2009}{#1} }
\newcommand{\etal}{\MakeLowercase{\textit{et al. }}} % "et al."

\begin{document}
\title{Development of HPD Clusters for MAGIC-II}

\author{\IEEEauthorblockN{R.~Orito\IEEEauthorrefmark{1},
			  E.~Bernardini\IEEEauthorrefmark{2},
                          D.~Bose\IEEEauthorrefmark{3},
                          A.~Dettlaff\IEEEauthorrefmark{1},
                          D.~Fink\IEEEauthorrefmark{1},
			  V.~Fonseca\IEEEauthorrefmark{3},
			  M.~Hayashida\IEEEauthorrefmark{1}$^{,1}$, \\
			  J.~Hose\IEEEauthorrefmark{1},
			  E.~Lorenz\IEEEauthorrefmark{1}\IEEEauthorrefmark{4},
			  K.~Mannheim\IEEEauthorrefmark{5},
			  R.~Mirzoyan\IEEEauthorrefmark{1},
			  O.~Reimann\IEEEauthorrefmark{1},
			  T.~Y.~Saito\IEEEauthorrefmark{1},\\
			  T.~Schweizer\IEEEauthorrefmark{1},
			  M.~Shayduk\IEEEauthorrefmark{1} and
			  M.~Teshima\IEEEauthorrefmark{1}, 
			  for the MAGIC collaboration
  }
                            \\
\IEEEauthorblockA{\IEEEauthorrefmark{1} Max-Planck-Institute fuer Physik, 
D-80805 Muenchen, Germany}
\IEEEauthorblockA{\IEEEauthorrefmark{2} DESY Deutsches Electr.-Synchrotron, D-15738
Zeuthen, Germany }
\IEEEauthorblockA{\IEEEauthorrefmark{3} Universidad Complutense, E-28040 Madrid, Spain }
\IEEEauthorblockA{\IEEEauthorrefmark{4} ETH, Zurich, CH-8093 Switzerland}
\IEEEauthorblockA{\IEEEauthorrefmark{5} Universitaet Wuerzburg, D-97074 Wuerzburg, Germany}}

\shorttitle{R.~Orito \etal MAGIC-II HPD CLUSTER}
\maketitle

\begin{abstract}
MAGIC-II is the second imaging atmospheric Cherenkov telescope
of MAGIC which is located on Canary island of La Palma.
%of the MAGIC observatory, which has recently been
%inaugurated on Canary island of La Palma.
We are currently developing a new camera based on clusters 
of hybrid photon detectors (HPD) for the upgrade of MAGIC-II.
The photon detectors feature a GaAsP photocathode and an 
avalanche diode as electron bombarded anodes with internal gain,
and were supplied by Hamamatsu Photonics K.K. (R9792U-40).
The HPD camera with high quantum efficiency will increase 
the MAGIC-II sensitivity and lower the energy threshold. 
The basic performance of the HPDs has been measured and a 
prototype of an HPD cluster has been developed to be 
mounted on MAGIC-II. Here we report on the status of
the HPD cluster and the project of 
eventually using HPD clusters in the central area 
of the MAGIC-II camera.
\end{abstract}

 
\section{Introduction}
MAGIC\cite{bai} is an observatory of two imaging atmospheric Cherenkov 
telescopes (IACT) for ground-based very high energy gamma-ray astronomy. 
It is located on Canary island 
of La Palma(28.75$^{\circ}$N, 17.86$^{\circ}$W, 2225~m a.s.l.). 
The first telescope, MAGIC-I is in operation since 2004. 
Stereoscopic observation with the second telescope, MAGIC-II, 
is planned to start in 2009\cite{thomas}. 
Using the two telescopes with mirrors of 17~m diameter, 
the sensitivity of MAGIC is improved by at least a 
factor of 2\cite{pierre}. 
The current camera of MAGIC-II consists of 1039 pixels of 
Hamamatsu R10408 photomultipliers (PMT) with a super-bialkali 
photocathode\cite{daniela}. 
To achieve a higher gamma-ray sensitivity with MAGIC-II, 
we plan to use a new camera consisting of photon detectors 
with higher performance\cite{bai}\cite{bar}\cite{razmik}\cite{echart}. 
One of the promising candidates for the new photon detector 
is a hybrid photon detector (HPD) with high quantum efficiency (QE). 
For MAGIC-II Hamamatsu Photonics K.K. has recently 
developed Hamamatsu R9792U-40 HPD with a GaAsP photocathode 
in collaboration with MPI Munich.
We have measured the basic performance of the 
R9792U-40\cite{hayashida}\cite{saito}, 
and developed a cluster to be mounted on the MAGIC-II 
camera for gaining experience in operation on IACTs. 
Here we report on the studies with a single HPD
and the prototype of the cluster to be tested on MAGIC-II.
We also report on the plan to build a new camera with HPDs
for MAGIC-II.

\section{Performance of the Hamamatsu R9792U-40 HPD}

 \begin{figure}[!t]
  \centering
  \includegraphics[width=2.8in]{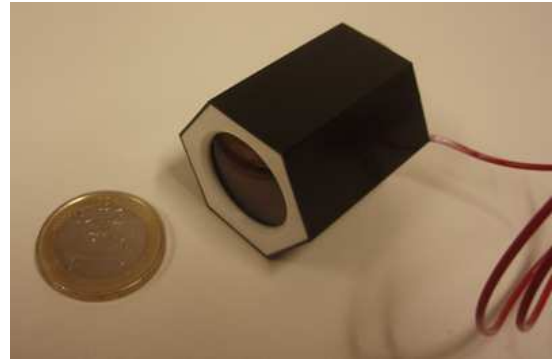}
  \caption{Photograph of Hamamatsu R9792U-40 HPD.}
  \label{hpd1}
 \end{figure}

\begin{figure}[!t]
  \centering
  \includegraphics[width=2.7in]{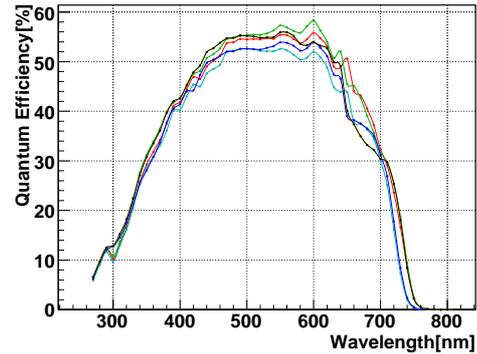}
  \caption{QE of the HPD R9792U-40 as a function of the wavelength.}
  \label{QE}
 \end{figure}

\begin{figure}[!t]
  \centering
  \includegraphics[width=2.7in]{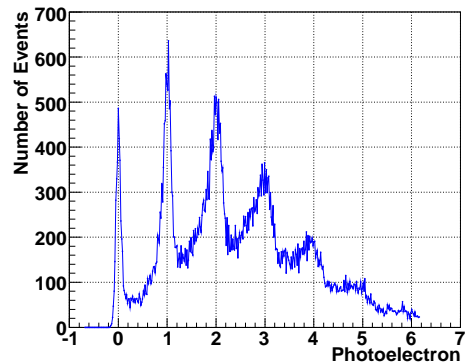}
  \caption{Single photoelectron resolution measured with HPD R9792U-40.}
  \label{peres}
 \end{figure}

\begin{figure}[!t]
  \centering
  \includegraphics[width=2.7in]{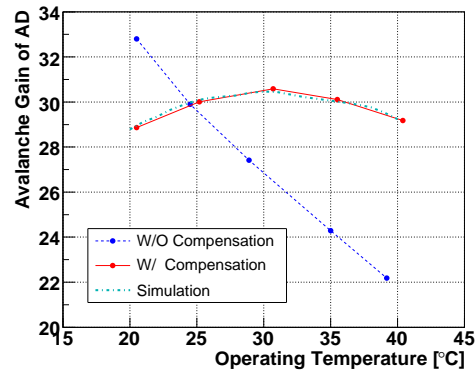}
  \caption{Temperature compensation of HPD R9792U-40.}
  \label{temp}
 \end{figure}

\begin{figure}[!t]
  \centering
  \includegraphics[width=2.7in]{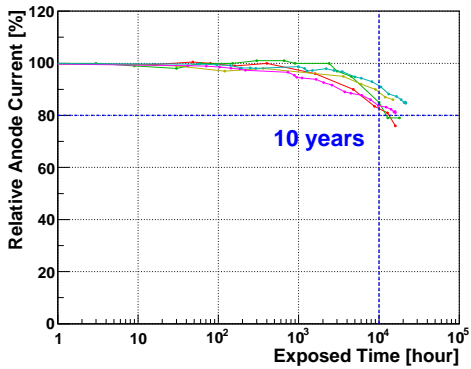}
  \caption{Lifetime estimation of HPD R9792U-40.}
  \label{life}
 \end{figure}

\begin{figure}[!t]
  \centering
  \includegraphics[width=3.2in]{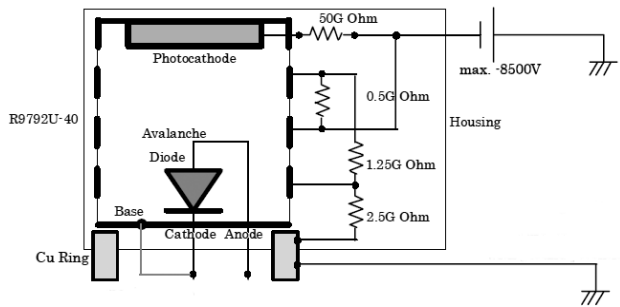}
  \caption{Protection circuit implemented to HPD R9792U-40.}
  \label{protect}
 \end{figure}

Figure~\ref{hpd1} shows a photograph of the Hamamatsu R9792U-40 
HPD of a compact, hexagonal structure. 
The Hamamatsu R9792U-40 HPD consists of a GaAsP photocathode
 and a 3~mm diameter avalanche diode (AD) acting as an 
 electron bombarded anode with additional internal gain. 
The diameter of the photocathode with uniform sensitivity is $\sim$18~mm. 
The bombardment and avalanche gain of the HPD R9792U-40 are $\sim$1550 
and $\sim$50 respectively, when biasing the photocathode 
at -8~kV and the AD at 400~V.
There are several important advantages of using the HPD R9792U-40 for IACTs. 
The first one is the high quantum efficiency of the GaAsP photocathode, 
which is above 50~$\%$ at 500~nm as shown in Fig.~\ref{QE}. 
The QE in UV region (300 - 400~nm) is rather low but can 
be increased by a wavelength shifter coating, 
which is a mixture of 0.03~g POPOP and 0.03~g butyl-PBD 
as a wavelength shifter, and 1.5~g of 
Paraloid B72 (an acrylic lacquer base) as a binder. 
This mixture is dissolved in Toluene (or similar solvent 
of medium evaporation speed at room temperature) and applied 
as a thin layer onto the window. 
After the evaporation of the Toluene, a hard layer is formed. 
In addition to the high QE, the photoelectron collection efficiency 
of the HPD R9792U-40 is almost 100~$\%$ because of the electrode structure 
and very high acceleration voltage. 
By using the HPD R9792U-40, we should obtain twice 
as many photoelectrons from atmospheric Cherenkov light 
as compared with the MAGIC-I camera PMTs. 
Figure~\ref{peres} shows the photoelectron resolution 
measured with an HPD R9792U-40. 
The pulse width of a narrow light flash illuminating the 
HPD R9792U-40 is 2.1~nsec (FWHM), which is about the typical 
time spread of Cherenkov light from low energy gamma showers, 
i.e. it is fast enough to enable proper 
operation under the night sky background (NSB) maximizing 
signal-to-noise ratio. 
The lower afterpulse rate of HPD R9792U-40 
compared to standard PMTs is also a key feature to minimize fake triggers.
The AD shows quite some gain dependence on temperature changes. 
Therefore we have implemented a thermistor on the copper ring, 
which is thermally coupled to the AD and corrects the bias 
voltage thus successfully compensates the temperature dependence 
of the AD gain, reducing it to an acceptable 
stability of $\sim$0.3$\%$/$^\circ$C in 
the temperature range between 25 to 35$^\circ$C. 
Figure~\ref{temp} shows both the temperature-compensated and 
uncompensated AD gain as a function of the operating temperature. 
The aging of the photosensors due to the rather NSB is also an
important issue. We have estimated the expected lifetime of the GaAsP photocathode 
to be more than 10~years operation under the dark night condition. 
Figure~\ref{life} represents the result of the lifetime estimation 
of HPD R9792U-40 from an accelerated aging test exposing the 
HPD to continuous light, which is 30$\sim$50 higher brighter than the nominal NSB. 
Here we define the lifetime as the period in which the QE degrades by 20~$\%$. 
A long lifetime was realized by suppressing ion feedback 
in the HPD R9702U-40 by a very high vacuum and a special 
ion deflector close to the AD\cite{ion}. 
We have also implemented a protection circuit 
against strong light like from car headlights during observation,
as shown in Fig.~\ref{protect}. 
All the basic performance parameters of HPD R9792U-40 seem to 
satisfy the requirements to operate such photon detectors on IACTs.

\section{Development of the HPD Cluster}
\begin{figure}[!t]
  \centering
 \includegraphics[width=3.2in]{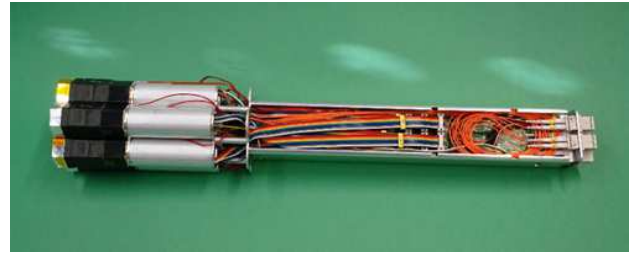}
  \caption{Side view of the HPD cluster for MAGIC-II.}
  \label{pro}
 \end{figure}

\begin{figure}[!t]
  \centering
  \includegraphics[width=3.2in]{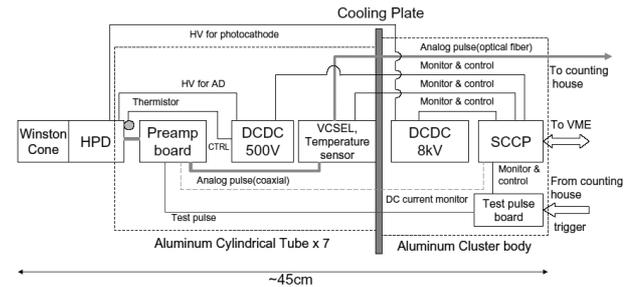}
  \caption{Schematic view of the HPD cluster module for MAGIC-II.}
  \label{sche}
 \end{figure}

\begin{figure}[!t]
  \centering
  \includegraphics[width=1.45in]{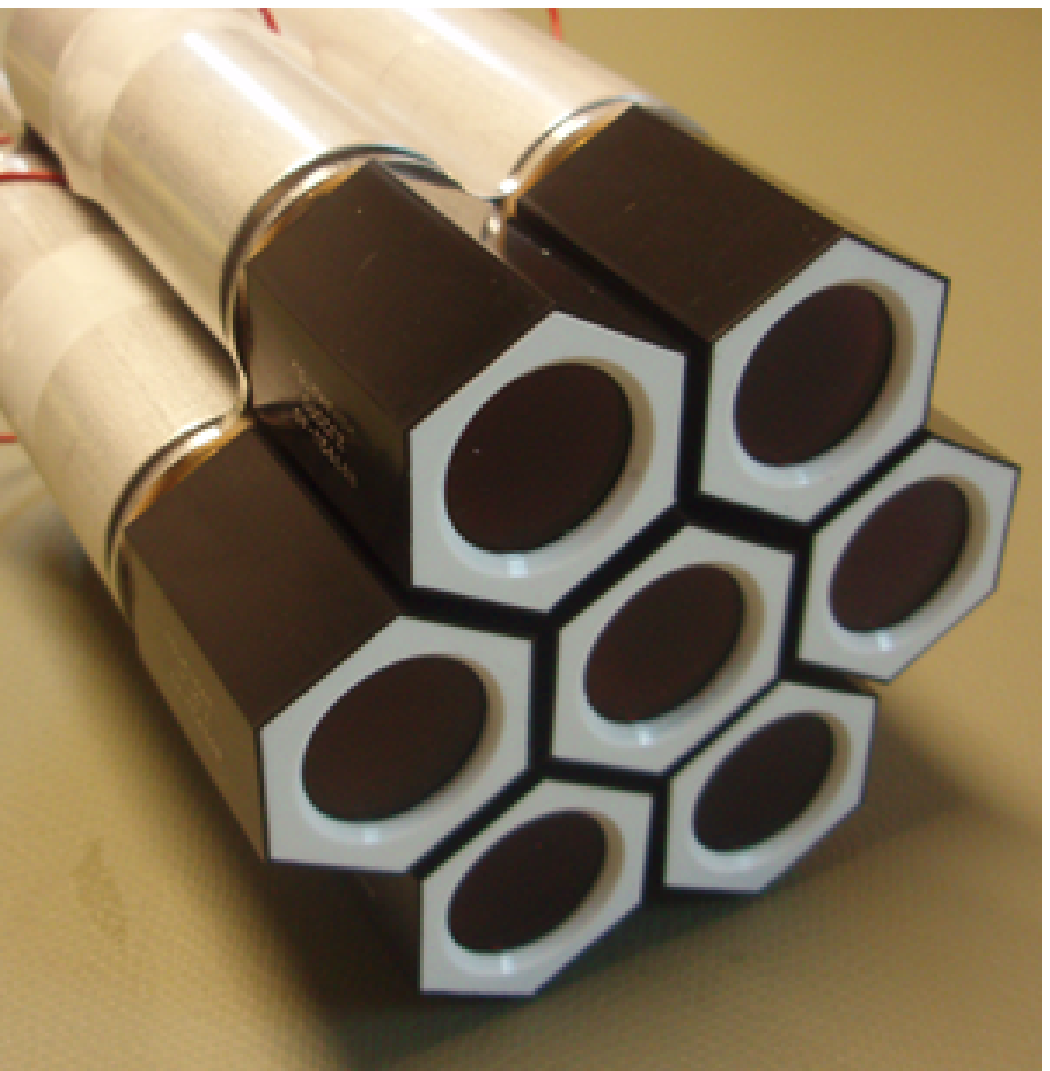}
  \includegraphics[width=1.65in]{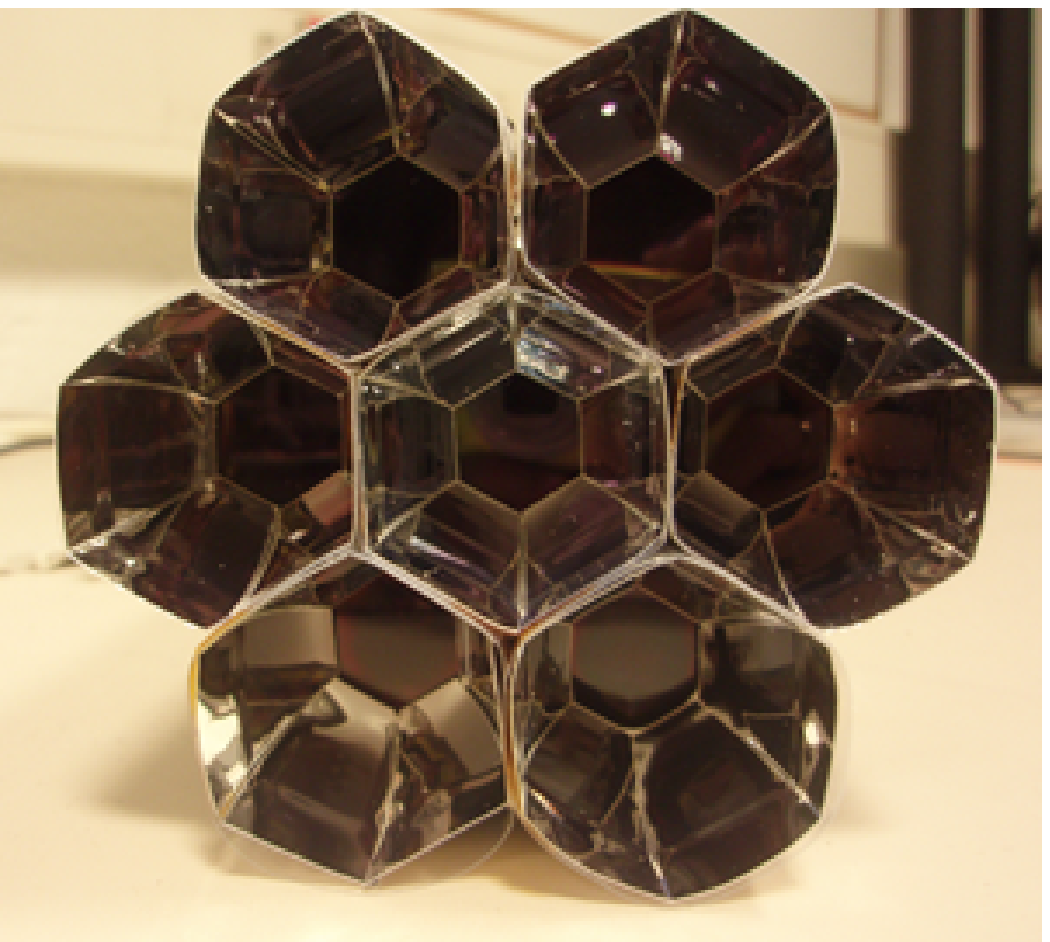}
  \caption{Front view of the HPD cluster for MAGIC-II without (left) 
and with light catchers (right). 
The light cathcers are made using the non-conductive UV-reflective 
dielectric film.}
  \label{wc}
 \end{figure}

To evaluate the HPD performance of the MAGIC-II camera, 
we have developed an HPD cluster module consisting of seven HPDs. 
Figure~\ref{pro} shows a prototype of the HPD cluster module. 
The geometry and interface of the HPD cluster are compatible 
to the current PMT cluster\cite{daniela}, i.e. plug exchangeable 
for possible future replacements. The HPD cluster 
module consists of an HPD, thermistor, preamp, VCSEL chip, DCDC-converters for 
the AD and photocathode, light catchers and slow
control cluster processor (SCCP). 
Figure~\ref{sche} shows a schematic view of the HPD cluster module. 
The AC coupled signal from the HPD is fed to a preamplifier with 
25~dB amplification and 700~MHz bandwidth (SIRENZA, SGA-5586). 
A test pulse system is also implemented to check the entire signal 
chain including the digitization and the readout. 
The amplified pulse signal is fed to the VCSEL chip (Avalon AVAP-850SM), 
which is the acronym for vertical-cavity surface-emitting laser 
and converts the electric signal to 850~nm wavelength optical signal. 
The optical signal is transferred to the counting house with 160 m multi-mode 
optical fibers. The VCSEL board is firmly mounted to the cooling plate 
of the camera for proper temperature stabilization. 
The SCCP\cite{haberer} controls the DCDC converters (SDS APD
Series for 400~V and SDS HPDB5802300 for -8~kV), 
and the current of the VCSEL chip. 
The DC current due to the steady flux of the NSB in the AD of the HPD, 
the current and temperature of the VCSEL, are also monitored via SCCP. 
The SCCP consists of a microcontroller chip (AMTEL AT89C51ID2), 
a 12bit ADC, and DAC. Each SCCP is connected to a VME board located 
in the camera with LAN cable (RJ45 connector). 
The VME board in the camera is connected to the PC in 
the counting house with an optical PCI to VME link (Str\"{u}ck SIS3100). 
The thermistor, preamp, VCSEL chip and 500~V DCDC converters 
are placed in seven cylindrical aluminum tubes used for EMI shielding. 
The -8~kV DCDC converters are common for all 
seven HPDs and placed next to the SCCP in the aluminum 
cluster body with careful high voltage insulation. 
``Light catchers'', similar to Winston cones, are used to 
provide nearly 100$\%$ light concentration onto 
the photocathodes and shield against large angle stray light. 
Figure~\ref{wc} shows a photograph of the HPD cluster  
and the light catchers. 
For using light catchers in contact with the HPD glass 
window with high voltage on the inner side, 
we have developed a novel non-conductive UV-reflective 
dielectric film in collaboration with Fraunhofer Institute\cite{wc}. 
As base material we use the 3M Vikuiti ESR2 foil with nearly 100~$\%$ 
reflectivity between 385 and 750~nm. For extending the high reflectivity 
down to 300~nm, the foil is overcoated with 40 alternative layers of 
SiO$_{2}$ and HfO$_{2}$. Currently, a mean reflectivity of 
95~$\%$ has been achieved when averaging from 300 to 750~nm. 
The film for the light catcher is cut by laser and 
assembled to a hexagonal shape using special tools. 
As in the current MAGIC-II camera the amplified signals 
are back-converted by VCSELs to optical signals, 
which are sent by optical fibers to the receiver\cite{rec} in 
the counting house and fed to the data acquisition system consisting of
Domino Ring Sampler\cite{Domino}. The Cherenkov signals
are digitized with 2GSample/s.

\section{Future Plan}
%At first
As the first stage,
we will mount six HPD cluster modules (42 pixels) on 
the corners of the MAGIC-II camera. Once a satisfactory 
performance of the HPD cluster is confirmed, more clusters 
will be mass-produced and the central 61 clusters with 
classical PMTs (427 pixels) of the MAGIC-II camera will be 
replaced in the near future by HPD ones.
%As the first stage, we mount six HPD 
%clusters(42 pixels) on the corners of the MAGIC-II camera.
%Once satisfactory HPD cluster performance is confirmed,
%the cluster will be mass-produced and the central 61 clusters(427 pixels) 
%of the MAGIC-II camera will be replaced in the near future.
%If the HPD clusters are well performed, the cluster is
%mass-produced and  the central 61 clusters(427 pixels) 
%of the MAGIC-II camera is replaced in near future.

\section{Conclusion}
A new camera of Hamamatsu R9792U-40 HPDs is presently under 
construction for future upgrades of the MAGIC-II telescope. 
The basic performance parameters of HPDs were tested and it has been 
found that they fulfill the requirements to be used on IACTs. 
We have developed an HPD cluster of seven HPDs with all the 
associated electronics. The cluster is plug-compatible to 
the current PMT clusters of the MAGIC-II camera. It is planned to 
replace the central region of the current MAGIC-II 
camera by HPD clusters in future to enhance the gamma-ray sensitivity. 
The status of the HPD development for MAGIC-II has been reported.
% in future no ichi.

\section*{Acknowledgements}
We would like to thank Hamamatsu Photonics K.K. for
developing the HPD designed for the MAGIC telescope.
The MAGIC project is supported by MPG
(Max-Planck-Society in Germany), BMBF
(Federal Ministry of Education and Research in
Germany), INFN (Italy), CICYT (Spain)
and IAC (Instituto de Astrophysica de Canarias).
This work is also supported by ETH Resarch Grant TH 34/043,
Polish MNiSzW Grant N N203 390834, and YIP of the
Helmholtz Germeinschaft.

\vspace*{0.5cm}
$^1$~Present address: Stanford Linear Accelerator Center, 2575 Sand Hill Road,
Menlo Park, CA94025, USA.


\begin{thebibliography}{99}
 \bibitem{bai} C.~Baixeras et al., ``Commissioning and first
tests of the MAGIC telescope'' Nucl.~Instr.~Meth.~A, 518 (2004) 188.
  \bibitem{thomas}  T.~Schweizer et al., ICRC2009 highlight talk, in these proceedings.
  \bibitem{pierre}  P.~Colin et al., ``Performance of the MAGIC telescopes in stereoscopic mode'', in these proceedings.
  \bibitem{daniela} D.~Borla Tridon et al., ``Performance of the Camera of MAGIC II Telescope'', in these proceedings.
 \bibitem{bar}J.A.~Barrio, et al., ``The MAGIC Telescope Design
Report'', MPI Institute Report, MPI-PhE/98-5 (March 1998).
 \bibitem{razmik} R.~Mirzoyan et al., ``An
evaluation of the new compact hybrid photodiodes
R7110U-07/40 from Hamamatsu in
high-speed light detection mode'',
Nucl.~Instr.~Meth.~A, 442 (2000) 140.
 \bibitem{echart} E.~Lorenz et al,
``Progress in the development of a high QE, red extended hybrid
photomultiplier for the second phase of
the MAGIC telescope'', Nucl.~Instr.~Meth.~A, 504 (2003) 280.
\bibitem{hayashida} M.~Hayashida et al.,
``Development of HPDs with an 18-mmdiameter
GaAsP photo cathode for the MAGIC-II'', 
Nucl.~Instr~.~Meth.~A, 567 (2006) 180.
  \bibitem{saito}   T.Y.~Saito et al., ``Very High QE HPDs with a GaAsP photocathode for the MAGIC Telecope Project'', to be appeared in Nucl.~Instr.~Meth.~A.
  \bibitem{ion} D.~Ferenc et al., ``Solution to the ion feedback problem in hybrid photon detectors and photomultiplier tubes'',  Nucl.~Instr~.~Meth.~A, 427 (1999) 518.
  \bibitem{haberer}  \url{http://www.mpp.mpg.de/~haberer/projects/MAGIC/magic_All.html}
  \bibitem{wc}  E.~Lorenz and M.~Shayduk, internal communication.
  \bibitem{rec}  J.M.~Illa et al., ``The receivers boards of the MAGIC-II Cherenkov telescope'', in these proceedings.
  \bibitem{Domino}  D.~Tescaro et al., ''The readout system of the MAGIC-II Cherenkov telescope'', in these proceedings.
\end{thebibliography}
\end{document}